\documentclass[a4paper]{jpconf}
\usepackage{physics}
\usepackage{graphicx}  
\graphicspath{{images/}{images2/}}  
\usepackage{wrapfig} 
\usepackage{subfigure} 
\usepackage{amsmath,amssymb}

\usepackage{cite} 
\usepackage{csquotes} 
\usepackage{graphicx}

\begin{document}
\title{Neutrino magnetic moments in low-energy neutrino scattering on condensed matter systems}

\author{Georgy Donchenko$^1$, Konstantin Kouzakov$^1$ \\and Alexander Studenikin$^{1,2}$}
\address{$^1$ Faculty of Physics, Lomonosov Moscow State University, Moscow 119991, Russia}
\address{$^2$ Joint Institute for Nuclear Research, Dubna 141980, Moscow Region, Russia}

\ead{dongosha@gmail.com}

\begin{abstract}{The cross sections of elastic neutrino scattering on electrons and nuclei in the regime of low-energy transfer are known to be very sensitive to neutrino electromagnetic properties. In particular, the magnetic moment of the neutrino can be effectively searched using liquid or solid detectors with a very low energy threshold. We present the formalism that incorporates the neutrino magnetic moment contribution in the theoretical treatment of the low-energy elastic neutrino scattering on a condensed-matter target. The concept of the dynamic structure factor is employed to describe the collective effects in the target. The differential cross section for tritium antineutrino scattering on the superfluid $^4$He is calculated numerically. We find that the neutrino magnetic moment of the order of $10^{-11}\mu_B$ strongly affects the cross section. Our results can be used in the search of neutrino magnetic moments in future low-energy neutrino scattering experiments with liquid or solid targets.
}\end{abstract}


\section{Introduction}
    Recently, it was proposed to study the coherent elastic neutrino-atom scattering~\cite{Cadeddu} using a superfluid $^4$He detector of light dark matter particles~\cite{Maris2017}. The detector is sensitive to low-energy signals on the order of $\sim$1 meV, and this feature can be effectively exploited for searching the neutrino magnetic moment~\cite{RMP2015}. In the present work we develop a formalism which takes into account the neutrino magnetic moment effects in the process of low-energy neutrino scattering on atoms in a liquid or a solid target. In our formalism we also account for the collective effects due to the interaction between the atoms in the target. 
    

%
\section{Theoretical formulation}
      In the elastic neutrino-atom collision the atom recoils as a whole and its internal state remains unchanged. We will assume that the neutrino energy satisfies the conditions $ E_\nu \ll m $ and $ E_\nu \ll 1 / R_{\rm nucl} $, where $ m $ is the atomic mass and $ R_{\rm nucl} $ is the radius of the atomic nucleus. According to the energy conservation, the kinetic energy of the recoil atom is equal to the energy transfer
        \begin{equation}\label{eq: relations}
        T\leq\frac{2E_\nu^2}{m}\ll E_\nu.
        \end{equation}
		%

        The differential cross section for elastic neutrino-atom scattering is given by the following expression~\cite{Gaponov}:	        %
        \begin{equation}
					\label{cr_sec_atom}
                      \left( \frac{d\sigma}{dT} \right)_{\rm atom} = \int\limits_0^{\infty} d{q}^2 \left[ \Sigma^{(w)} + \Sigma^{(\mu)} \right]\delta\left(T-\dfrac{q^2}{2m}\right).
        \end{equation}
        Here
		\begin{align}
		        \label{SigmaW}
		                 \Sigma^{(w)}& = \frac{G_F^2 }{2\pi}  \left[ C_V^2 \left(1 - \frac{{q}^2}{4 E_\nu^2} \right) + C_A^2 \left(1 + \frac{{q}^2}{4 E_\nu^2} \right) \right],\\
				\label{C_V}
            		    C_V &= Z \left(\frac{1}{2} - 2 \sin^2 \theta_W \right) - \frac{1}{2} N + Z \left( \pm\frac{1}{2} + 2 \sin^2 \theta_W \right) F_{\rm el} ( q^2 ),\\
            	\label{C_A}	    
    					C_A^2 &= \frac{g_A^2}{4} \left[ (Z_{+} - Z_{-} ) - (N_{+} - N_{-}) \right]^2 + \frac{1}{4}\sum_{n=1,2,\dots}\sum_{l=0}^{n-1}\left|\left(  L^{nl}_{+} - L^{nl}_{-} \right) F_{\rm el}^{nl}(q^2)\right|^2,
            \end{align}
        where $q$ is the momentum transfer, with $q^2=2mT$, the plus stands for $\nu_e$ and $\bar{\nu}_e$, and the minus for all other neutrino species), and $Z$ ($N$) is the number of protons (neutrons). $F_{\rm el}(q^2)$ ($F_{\rm el}^{nl}(q^2)$) is the Fourier transform of the electron density (the electron density in the $nl$ atomic orbital), $g_A=1.25$, and $Z_\pm$ and $N_\pm$ ($L^{nl}_\pm$) are the numbers of protons and neutrons (electrons) with spin parallel ($+$) or antiparallel ($-$) to the nucleus spin (the total electron spin).
          The neutrino magnetic moment contribution is~\cite{Cadeddu}
                \begin{equation}
					\label{SigmaMu}
                      \Sigma^{(\mu)} = \frac{\pi\alpha^2 Z^2}{m_e^2} |\mu_\nu|^2 \left(\frac{2m}{q^2} - \frac{1}{E_\nu} \right) F_{\rm scr}^2,
        \end{equation}
        where $\mu_\nu$ 
        is the neutrino dipole magnetic moment in units of $\mu_B$ 
        and $F_{\rm scr} =1 - F_{\rm el}(q^2)$.

%
%
    Consider now low-energy neutrino scattering by the system of  $ \mathcal{N} $ interacting atoms (a liquid or a solid target). The energy $ T $ transferred by the neutrino to one of the atoms can be redistributed between the atoms in the system due to their interaction. The initial (final) state of the system and its energy are $ | i \rangle $ ($ | f \rangle $) and $ E_i $ ($ E_f $). For a single-atom system ($ \mathcal{N} = 1 $) we have
    \begin{equation}
    \label{d3s_if}
    \frac{d\sigma_{i\to f}}{dT d{q}^2d{\varphi_q}}= \frac{1}{2 \pi}\left[ \Sigma^{(w)} + \Sigma^{(\mu)} \right] \delta(T-E_f+E_i),
    \end{equation}
	where $ \varphi_q $ is the azimuthal angle of the momentum transfer ${\vec q} $ (the $ z $ axis is directed along the initial neutrino momentum), $ E_f-E_i ={q^2} / 2m $. The result~(\ref{d3s_if}) is generelized to the case of $ \mathcal{N} $ atoms by means of the following substitutions:
    \begin{equation}
    \label{C_N}
    C_{V,A}^2\to\left|C_{V,A}^{(\mathcal{N})}\right|^2=\left|\langle f|\sum_{j=1}^{\mathcal{N}}e^{i\vec{q}\vec{R}_j}C_{V,A}|i\rangle\right|^2, \qquad F_{\rm scr}^2\to\left|F_{\rm scr}^{(\mathcal{N})}\right|^2=\left|\langle f|\sum_{j=1}^{\mathcal{N}}e^{i\vec{q}\vec{R}_j}F_{\rm scr}|i\rangle\right|^2,
    \end{equation}
    where $ \vec{R}_j $ denotes the position of the $ j $th atom. 
	Summing over all possible final states and averaging over the initial states, we find
	\begin{align}
	\label{cr_sec}
	\frac{d\sigma}{dT}=&\int\limits_0^{\infty} d{q}^2 \int\limits_0^{2\pi} d{\varphi_q}\,\frac{1}{2 \pi}\left[ \Sigma^{(w)} + \Sigma^{(\mu)} \right] S(T,{\vec q}).
	\end{align}
	Here we introduced the dynamic structure factor
    \begin{equation}
						\label{S(T,q)}
        S(T, \vec{q}) = \sum_{i,f} w_i \left|\langle f|\sum_{j=1}^{\mathcal{N}}e^{i\vec{q}\vec{R}_j}|i\rangle\right|^2\delta(T-E_f+E_i),
    \end{equation}
	where $ w_i $ is the statistical weight of the state $ | i \rangle $. 

     \begin{figure}[b]
        \includegraphics[width=0.8\linewidth]{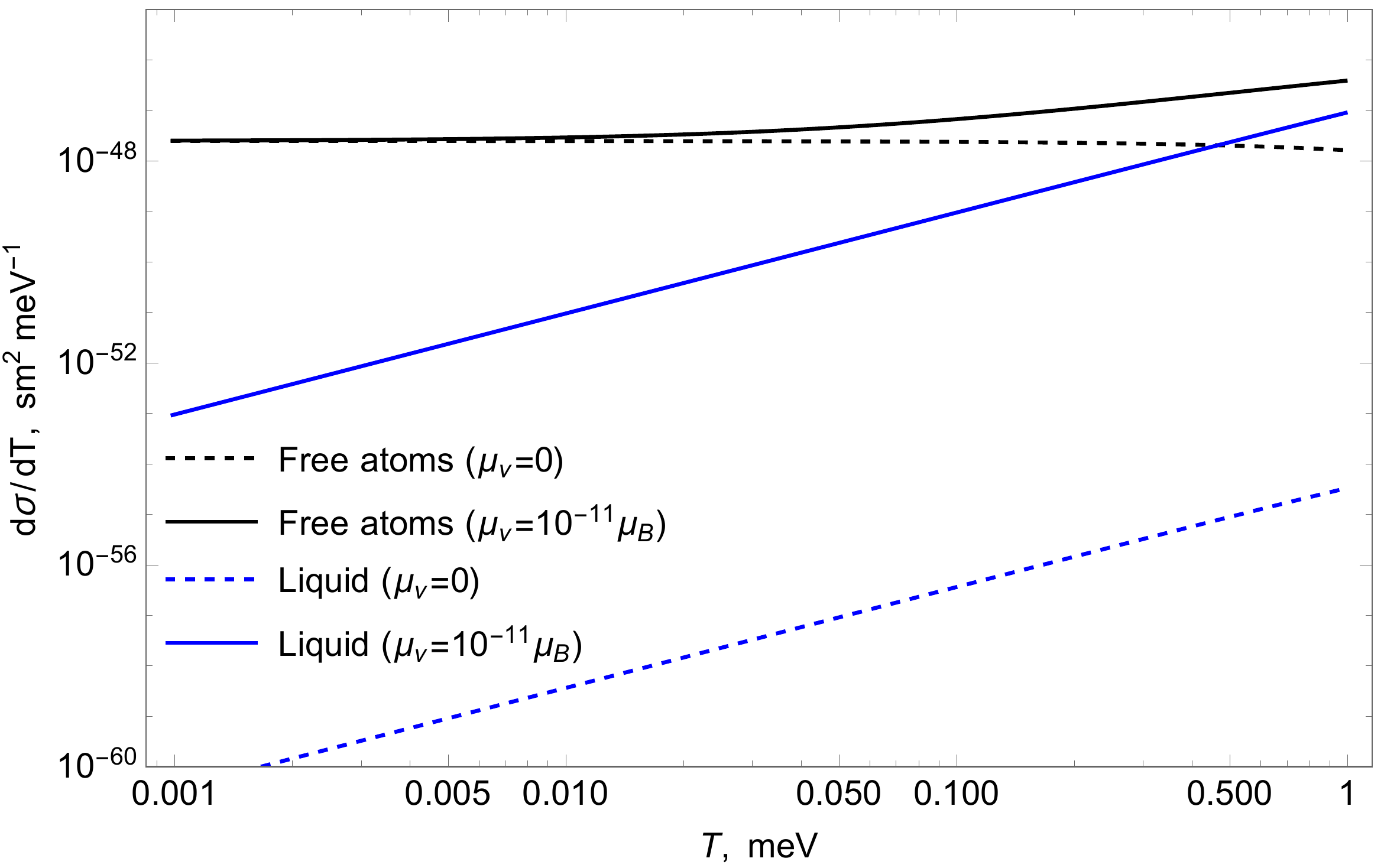}
               \centering
        \caption{The differential cross section for the tritium antineutrino scattering on the system of helium atoms at $E_\nu=10$~keV normalized to the number of atoms $\mathcal{N}$.} \label{fig:cs}
        \centering
    \end{figure}
     \begin{figure}[b]
        \includegraphics[width=0.8\linewidth]{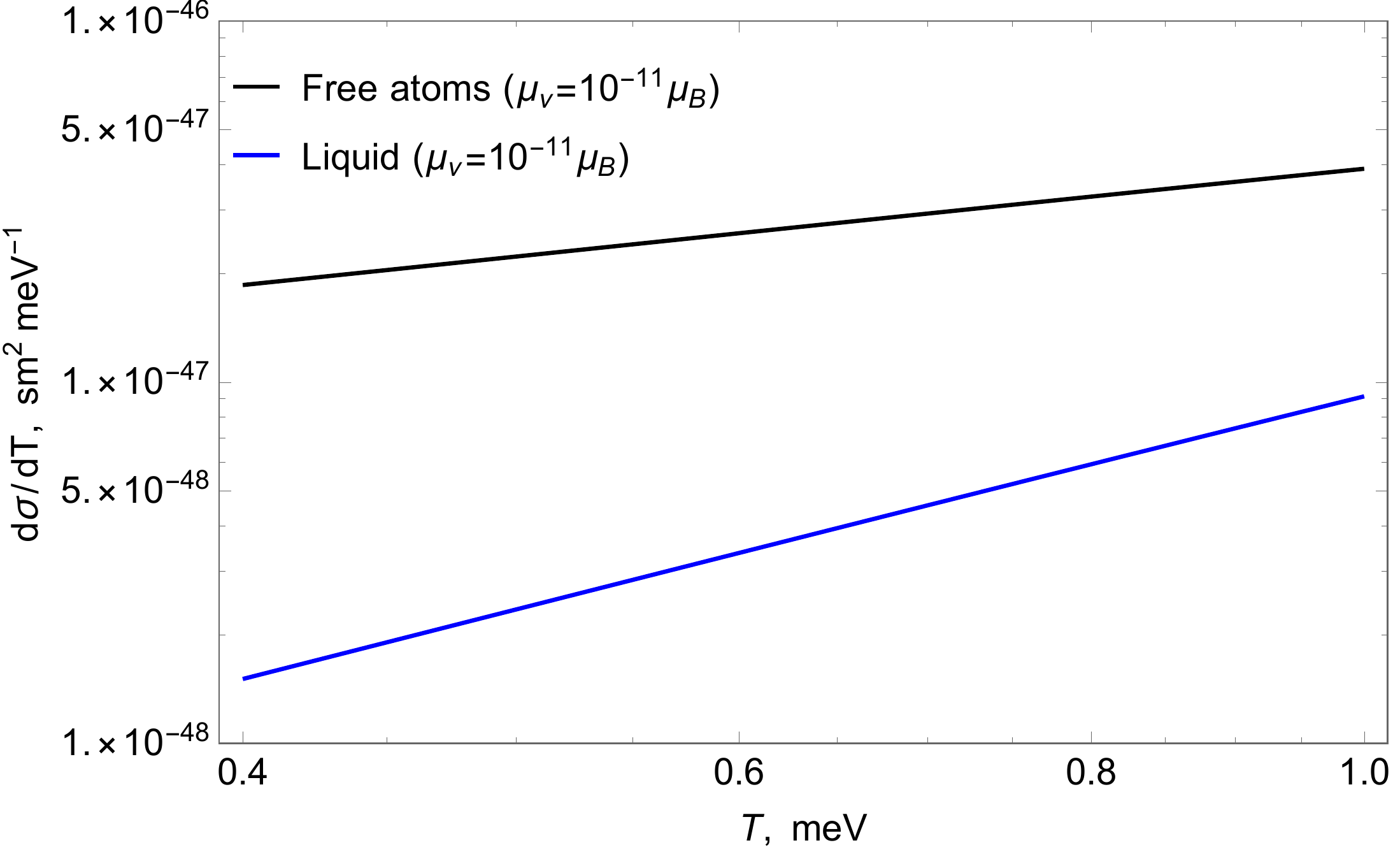}
               \centering
        \caption{The same as in fig.~\ref{fig:cs}, but only for a nonzero neutrino magnetic and energy-transfer values 0.4 meV$\leq T\leq$1 meV. 
        } 
        \label{fig:cs_o_l}
        \centering
    \end{figure}
    
\section{Numerical results}
   We consider tritium antineutrino scattering on a superfluid $^4$He target to illustrate the developed formalism and to point out the roles of the neutrino magnetic moment and the collective effects.
    The dynamic structure factor for the energy-transfer values $T\lesssim1$ meV can be approximated as $ S(T, \vec{q}) = \langle\rho_{\vec q}\rho_{\vec q}^+\rangle\delta(T-uq)$~\cite{Kvasnikov}, where $u$ is the sound velocity in superfluid $^4$He. 
    For the cross section~(\ref{cr_sec}) we obtain
	\begin{align}
	\label{cr_sec_phonon}
	\frac{d\sigma}{dT}=\mathcal{N} \left[ C_V^2\,\frac{G_F^2 T}{2 \pi m u^2 } \left(1 - \frac{T^2}{4u^2E_\nu^2} \right) + \frac{\pi \alpha^2 Z^2}{m_e^2} |\mu_\nu|^2 \left(\frac{2m u^2}{T^2} - \frac{1}{E_\nu} \right) F_{\rm scr}^2 \right] \bigg|_{q = \frac{T}{u}}.
	\end{align}

    At the same time for the system of  $ \mathcal{N} $ non-interacting helium atoms we have        
    \begin{equation}
				\label{cr_sec_atom_He}
                  \frac{d\sigma}{dT} = \mathcal{N}\left[C_V^2\,\frac{G_F^2 m}{\pi}\left(1 - \frac{mT}{2E_\nu^2} \right) + \frac{\pi \alpha^2 Z^2}{m_e^2} |\mu_\nu|^2 \left(\frac{1}{T} - \frac{1}{E_\nu} \right) F_{\rm scr}^2 \right] \bigg|_{q^2 = 2mT}.
    \end{equation}
    %
    
    In fig.~\ref{fig:cs} we present the cross sections in the cases of interacting (liquid) and non-interacting (free) helium atoms. The numerical calculations when $\mu_\nu= 0$ show that the cross section with account for the collective effects (phonon excitations) is suppressed by many orders of magnitude as compared to the case of free atoms. However, when the magnetic moment is nonzero, $\mu_\nu = 10^{-11} \mu_B$, the suppression is found to be much weaker: the cross sections for the liquid and for the free atoms are almost of the same order of magnitude in the energy range $0.5$\,meV$\lesssim T\lesssim1$\,meV (see Fig.~\ref{fig:cs_o_l}).

\section{Summary and conclusions}
    We derived the differential cross section of elastic neutrino scattering on a condensed matter system in the regime of low-energy transfer, taking neutrino magnetic moments into account. We numerically calculated the cross section in the case of tritium antineutrino scattering on the liquid $^4$He in the supefluid phase. The results show that introducing nonzero neutrino magnetic moments boosts the cross section by a factor of $10^7$, making it close to the case of free atoms. This finding can be important for future neutrino experiments searching for neutrino electromagnetic properties~\cite{RMP2015} with detectors based on the liquid $^4$He and other materials (e.g., diamond~\cite{kurinsky2019} and graphene~\cite{Hochberg2017}). 
    

\ack
    This research has been supported by the Interdisciplinary Scientific and Educational School of Moscow University ``Fundamental and Applied Space Research'' and also by the Russian Foundation for Basic Research under Grant No. 20-52-53022-GFEN-a. The work of GD is supported by the BASIS Foundation No. 20-2-9-9-1.

\end{document}